\documentclass[preprint,12pt]{article}

\usepackage[english]{babel}

\usepackage{amsmath, amsfonts, amssymb}

\usepackage{graphicx}  
\usepackage{subfigure}
\usepackage{dcolumn}   
\usepackage{bm}        
\usepackage{multirow}
\usepackage{mathrsfs}
\usepackage{xcolor}
\usepackage{braket}
\usepackage{array}

\usepackage{lscape}

\begin{document}
\title{A complex Gaussian representation of continuum wavefunctions respectful of their asymptotic behaviour}
\author{Stéphanie Laure Egome Nana, Arnaud Leclerc  \\
 and Lorenzo Ugo Ancarani \\
Universit\'e de Lorraine, CNRS, \\
Laboratoire de Physique et Chimie Th\'eoriques, F-57000 Metz, France.}

\maketitle

\begin{abstract}


Complex Gaussian basis sets are optimized to accurately represent continuum radial wavefunctions over the whole space.
First, attention is put on the technical ability of the optimization method to get more flexible series of Gaussian exponents, in order to improve the accuracy of the fitting approach.
Second, an indirect fitting method is proposed, allowing for the oscillatory behaviour of continuum functions to be conserved up to infinity as a factorized asymptotic function, while the Gaussian representation is applied to some appropriately defined distortion factor with limited spatial extension. As an illustration, the method is applied to radial Coulomb functions with realistic energy parameters. We also show that the indirect fitting approach keeps the advantageous analytical structure of typical one-electron transition integrals occurring in molecular ionization applications.

\end{abstract}

\section{Introduction \label{introduction} }

Gaussian basis sets are of common use in quantum chemistry. Their mathematical properties are nowadays fully exploited to perform efficient molecular bound state calculations.
There are two main advantages of using Gaussian basis functions:
all the electronic transition integrals can be expressed in closed form and
this allows for simplification of multicentric integrals in the case of molecular problems (see e.g. \cite{hill_2013,gill1994molecular,shavitt1963}).
These practical advantages have motivated attempts to extend the use of Gaussian-type orbitals (GTOs) to   atomic and molecular phenomena involving \emph{continuum} states, with application to ionization processes or high harmonic generation~\cite{kaufmann_1989,Nestmann_1990,Faure_2002,fiori_2012, mavsin_2014,marante_2014,zhu_2021,coccia_2021, wozniak_2021,witzorky_2021,gao_2021,morassut_2022}.
In such applications, the quality of the physical results (e.g. ionization cross sections) relies also on the accuracy of the underlying strategy used to represent the continuum states.

Contrary to bound states, continuum wavefunctions intrinsically oscillate up to infinity while GTOs are monotonically decreasing and localized functions.
In the present paper, we focus on \emph{complex} Gaussian-Type orbitals (cGTOs), i.e. GTOs with complex exponents which are actually oscillating functions with Gaussian envelopes, more likely to efficiently represent continuum functions.
The general idea is to approximate a set of $p_{\text{max}}$ given radial oscillating functions $u_p(r)$, associated with a given continuum state, by a linear combination of $N$ cGTOs,
\begin{equation}
	u_{p}(r) \approx u_{p}^G(r) \equiv  \sum_{s=1}^{N}
		\left[ c_s \right]_{p}  e^{-\left[ \alpha_s \right]_{p} \, r^2}, \quad
		 p=1,\dots,p_{\text{max}}.
	\label{eq:cG_def}
\end{equation}
The Gaussian exponents are complex, $\alpha_s = \text{Re}(\alpha_s) + \imath \text{Im}(\alpha_s) $, with $\text{Re}(\alpha_s)>0$ and are optimized to minimize the difference between the original function and its Gaussian representation \cite{ammar_2020}.\footnote{Note that alternatively, a power of $r$ can been introduced before the Gaussian expansion to facilitate the convergence at short distances, i.e.
$ u_{p}(r) \approx r^{l+1}  \sum_{s=1}^{N}
		\left[ c_s \right]_{p}  e^{-\left[ \alpha_s \right]_{p} \, r^2}.
$
For simplicity of presentation we do not consider here this interesting variant.}
The choice of using such cGTO expansions to represent continuum functions is clearly motivated by the crucial advantage that all transition integrals are expressed in closed form, thus simpler to evaluate. This benefit is even more attractive in the case of multicentric integrals \cite{shavitt1963,ammar_2021_II}.
The cGTO method, inspired from previous work dealing with real Gaussians \cite{Nestmann_1990,Faure_2002,fiori_2012}, has recently produced interesting results in different applications to molecular photoionization \cite{ammar_2020,ammar_2021_I} and electron-impact ionization \cite{ammar_2024}.
However, based on our experience, we have come across two
 shortcomings
in the fitting strategy. In the present paper we would like to highlight them and propose ways to deal with them.

The first shortcoming seems at first rather technical.
When the set of complex Gaussian exponents is gradually modified during the optimization process, numerical instabilities can occur, requiring the specification of research bounds. This issue is dealt with by some optimization algorithm using successive search radii and research bounds for all variables.
Different orders of magnitude are obtained for the real part of the exponents, providing both tight and diffuse basis functions.
Due to the restricted search, large exponents can be somewhat artificially restrained because of global search radii and bounds, applied on the same footing to all the exponents to be optimized.
In section \ref{sec:dimensionless_search} we will give some details about this first  imperfection and solve the inconsistency by using dimensionless variables so that the same relative flexibility is allowed for all the exponents.

The second shortcoming is related to the asymptotics.
By construction, fitting an oscillating function with a finite sum of localized basis function such as cGTOs is only possible within a given, finite radial domain. Complex Gaussians oscillate, which is good, but still span a limited spatial extension.
The cGTO representation can only be accurate within some limited spatial box where the optimization algorithm is fed by numerical values of the target function.  Being aware of and mastering the way the approximate function behaves outside the fitting box is important
when used in subsequent physical applications.
In section \ref{sec:asymptotics}, we propose a simple method to make the cGTO representation of continuum radial functions consistent with their expected asymptotic oscillatory behaviour.
Since we apply a transformation to the function before the function is fitted, we call this method indirect fitting.
In order to illustrate the solutions proposed to deal with the two weaknesses, we consider the well known Coulomb radial functions. We finally investigate the consequences of our proposal on the calculation of one-electron integrals.

Atomic units are used throughout.

\section{Dimensionless optimization of the complex Gaussian exponents \label{sec:dimensionless_search} }

\subsection{Continuum wavefunction \label{ssec:continuumwf}}

In this paper, we focus on the radial part $u_{l,k}(r)$ of continuum wavefunctions,  solution of the following ordinary differential equation for a given angular momentum quantum number $l$:
\begin{equation}
	\left[ -\frac{1}{2} \frac{d^2}{dr^2} + \frac{l(l+1)}{2r^2}
	+ V(r)  \right] u_{l,k}(r)
	= \frac{k^2}{2} u_{l,k}(r)
	\text{,}
	\label{eq:fct_disto_DE}
\end{equation}
where $V(r)$ is an atomic or molecular central potential felt at radial position $r$ by an electron, escaping with momentum $k$.
For simplicity of the illustration, we consider here a pure Coulomb potential $V(r) = -z/r$, associated to an asymptotic charge $z=1$.
The Coulomb phase shift is then given by $\delta_l =\arg \left( \Gamma(l+1 + \imath \eta) \right)$
with the Sommerfeld parameter $\eta = -z/k$, and the radial functions are the regular Coulomb functions, defined as
\begin{equation}
\begin{aligned}
	u_{l,k}(r)
	=&
	(2k r)^{l+1} e^{-\frac{\pi \eta}{2}}
	\frac{\left|\Gamma\left(l+1+\imath \eta \right)\right|}
	{2\Gamma\left(2l+2 \right)} e^{\imath k r} \,
	\mathstrut_1 F_1 \left( l+1 + \imath \eta , 2l+2 ; - 2\imath k r \right)
	\text{,}
	\end{aligned}
	\label{eq:RegCoulFun}
\end{equation}
where $\mathstrut_1 F_1$ is the Kummer confluent hypergeometric function~\cite{bateman1953,gradshteyn2014}.
These radial functions are real functions.

\subsection{Summary of the general fitting strategy \label{ssec:fitting_strategy}}

We start with a summary of the fitting strategy previously described in ref. \cite{ammar_2020,ammar_2021_I}
which was based on previous works on real Gaussian optimizations \cite{Nestmann_1990,Faure_2002}.
We would like to represent one or a set of $u_p(r)$ continuum functions ($p$ stands for the collection of indices $l,k$) by a linear combination~\eqref{eq:cG_def} of $N$ cGTOs.
The optimal expansion is found by minimizing the objective function
\begin{equation}
\begin{aligned}
&\Xi
=
\sum_{p} \frac{\sum_{\kappa} | u_{p}(r_{\kappa}) - u_{p}^G(r_{\kappa}) |^2}
{\sum_{\kappa} | u_{p}(r_{\kappa}) |^2}
+ D(\text{Re}(\alpha_1),\dots,\text{Re}(\alpha_N)) \text{,}
\end{aligned}
\label{eq:objective}
\end{equation}
over some given radial grid $\{r_{\kappa}\}_{{\kappa}=1,\dots,{\kappa}_{max}}$.
The $\Xi$ function depends on $2N$ non-linear variables, $\{\text{Re}(\alpha_s),\text{Im}(\alpha_s) \}_{s=1,\dots,N}$
(the so-called exponents),
and $N \times p_{\text{max}}$ linear parameters
$\{[c_s]_{p}\}_{s=1,\dots,N,p=1,\dots,
		p_{max}}$
(the expansion coefficients).
In eq.~\eqref{eq:objective}, $D$ is a penalty function introduced to avoid the coalescence of two different exponents to the similar values (see details in \cite{ammar_2020}).

The optimization needs to be fed with a set of initial exponents $\{\alpha_s\}$.
Reasonable convergence can be reached by setting to zero the imaginary parts and taking the real parts between two bounds $\text{Re}(\alpha_1) = a$ and $\text{Re}(\alpha_N) = b > a$, following the distribution
\begin{equation}
\frac{\text{Re}(\alpha_{s+1})}{\text{Re}(\alpha_{s})} = \left(\frac{b}{a} \right)^{\frac{1}{N-1}}
\text{.}
\label{eq:initial_exp}
\end{equation}

The objective function $\Xi$ is minimized following a two step iterative algorithm:
(i) a least square optimization gives an approximation for the coefficients $\{c_s\}$
and
(ii)
the exponents $\{\alpha_s\}$ are optimized by an algorithm able to efficiently tackle non-linear minimization with selected constraints.
Steps (i) and (ii) are repeated until predefined convergence is obtained.

For step (ii) we use the Bound Optimization BY Quadratic Approximation (BOBYQA)~\cite{powell2009}.
It requires as an input the set of research bounds indicating that the exponents have to be optimized within some fixed intervals. The simplest choice applied in our previous works \cite{ammar_2020,ammar_2021_II,ammar_2024} is to  specify identical boundaries valid for all the exponents,
\begin{align}
& A_{\text{min}} < \text{Re}(\alpha_s) < A_{\text{max}} ,  \nonumber \\
& B_{\text{min}} < \text{Im}(\alpha_s) < B_{\text{max}}.
\end{align}
Besides those global limiting bounds, the optimization algorithm performs successive iterations using a ``trust region radius'' $\Delta_m$ so that during the $m^{th}$ iteration, the vectorial norm of the change in the optimized variables $ \parallel \vec{x} \parallel $ is not more than  $\Delta_m$ ($\vec{x}$ denotes the collection of all $\{ \text{Re}(\alpha_s), \text{Im}(\alpha_s) \}$ being explored).
The trust region radius is gradually decreased during the iteration from $\Delta_{beg}$ to $\Delta_{end}$. The optimization is stopped either if some epsilon value is reached by the objective function or, most of the time, when the last (smallest) trust region radius has been explored.

\subsection{Dimensionless search of the optimal exponents \label{ssec:dimensionless} }

The specification of common research bounds and common series of trust region radii is quite crude.
Allowing for different research bounds for different variables is likely to be a better choice, especially for the real parts $\text{Re}(\alpha_s)$ which span a rather large domain covering several orders of magnitude.
In what follows we will adopt and test a dimensionless version of the Gaussian exponent optimization in step (ii) of the optimization algorithm.

Instead of working directly with the complex numbers $\alpha_s$,
we define $2N$ dimensionless real optimization parameters $\{ \tilde{\alpha}^R_s,\tilde{\alpha}^I_s \}, s=1,\dots,N$ so that the Gaussian approximation to be optimized becomes
\begin{equation}
	u_{p}(r) \approx u_{p}^G(r) \equiv  \sum_{s=1}^{N}
		\left[ c_s \right]_{p}
		e^{-
		\left[
		\text{Re}(\alpha_s^{(0)}) \times \tilde{\alpha}^R_s
		+ \,
		\imath \,
		\text{Im}(\alpha_s^{(0)}) \times \tilde{\alpha}^I_s
		\right]_{p}
		\, r^2}.
	\label{eq:dimensionless_expansion}
\end{equation}
The $ \{ \alpha_s^{(0)} \}$ are fixed and correspond to the guess values before optimization.
The real parts of $ \{ \alpha_s^{(0)} \} $ can still be initialized following eq.~\eqref{eq:initial_exp}, while the imaginary parts are set to some fixed non-zero value.
The series of dimensionless numbers $\{ \tilde{\alpha}^R_s,\tilde{\alpha}^I_s \}$ are all initially set to $1$ and become the effective optimization variables.
Global fixed values of the research bounds are still given but they are now relative values, and the Gaussian exponents can vary proportionally to their initial values within those intervals:
\begin{align}
& \tilde{A}_{min} < \tilde{\alpha}^R_s < \tilde{A}_{max},  \nonumber  \\
& \tilde{B}_{min} < \tilde{\alpha}^I_s < \tilde{B}_{max}.
\end{align}
The real part parameter bounds must remain positive, $\tilde{A}_{min},\tilde{A}_{max} > 0$, and the imaginary part bounds are chosen so that each basis function oscillates with a reasonable 'frequency' over the main part of the associated Gaussian envelope.

In the same spirit, the successive trust region radii $\Delta_m$ are now associated with relative variable changes during a given iteration of the BOBYQA method.
In short, with the dimensionless search, short-range cGTOs remain short-range, diffuse cGTOs remain diffuse, but now all basis functions are allowed for a similar flexibility during the optimization.

\subsection{Numerical illustration with a set of Coulomb functions}

As an illustrative example we consider a set $\{u_{p}(r)\}_{p=1,\dots,8}$ of regular Coulomb functions~\eqref{eq:RegCoulFun} with angular quantum number $l= 0$
and momentum grid $k_{p}=0.5+0.25(p-1)$ a.u., $p=1,\dots,8$.
We optimize a set of $N=30$ cGTOs with complex exponents $\{ \alpha_1, \dots, \alpha_{30} \}$ over the radial interval $ R \in  ] 0, 30]$ a.u..
The number of cGTOs has been selected based on previous more detailed convergence studies, see \textit{e.g.} paragraph 3.4 of ref.~\cite{ammar_2023} and \cite{ammar_2020}.

Three different sets of optimization parameters are described in table \ref{tab:fitting_parametersABC}.
In optimization (A), the exponents are directly optimized and are constrained by common search radii  and by common absolute values of the research bounds.
For runs (B) and (C), the dimensionless search is applied following the explanations given in subsection \ref{ssec:dimensionless}.
The difference between run (B) and run (C) is only the value of the minimum research bound for $\text{Re}(\alpha_s)$, allowed to reach $10 \%$ or $1 \%$ of the initial value, respectively.

\begin{figure}

\begin{center}

\includegraphics[width=0.7\linewidth]{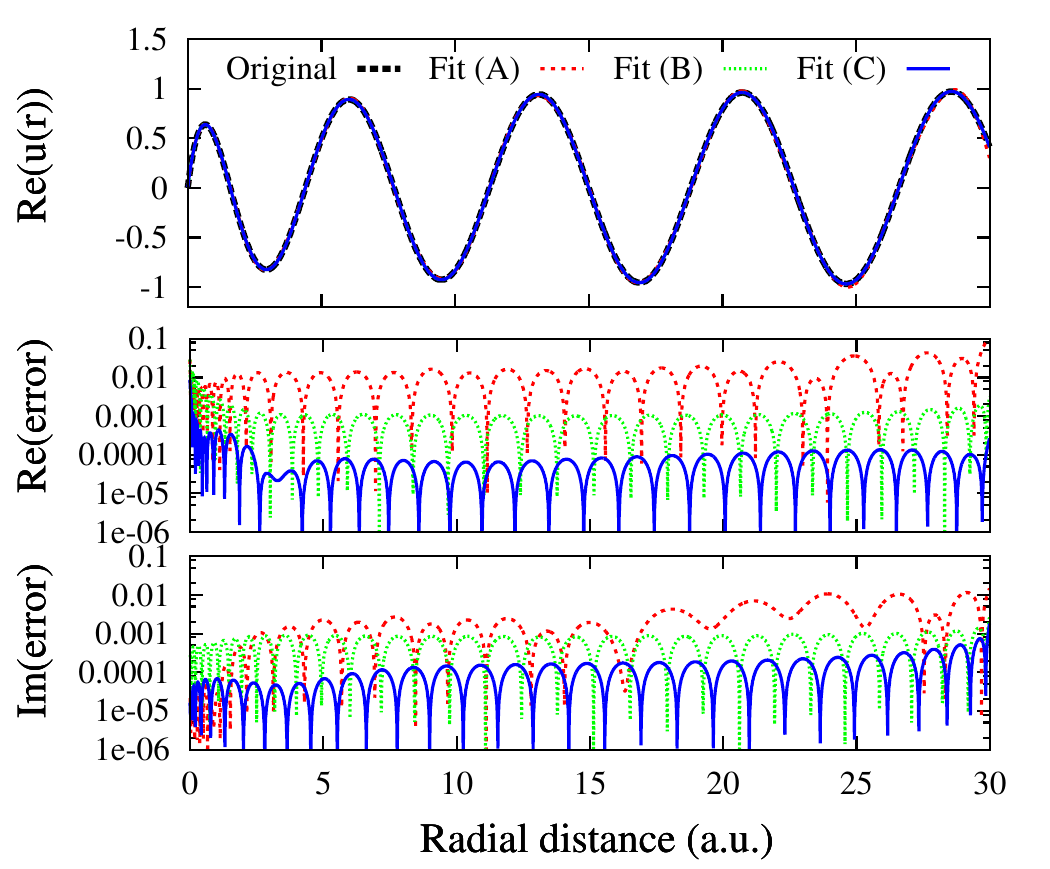}

\end{center}

\caption{ \label{fig1_fitting_errors}
Comparison of complex Gaussian expansions of the Coulomb radial function for $k=0.75$ a.u. and $l = 0$ represented with 30 cGTOs, obtained from three different sets of optimization parameters A, B, C as described in table \ref{tab:fitting_parametersABC}. The upper panel shows the function and its different Gaussian fits, the second and third panels show the real and imaginary parts of the error, respectively.}

\end{figure}

Fig. \ref{fig1_fitting_errors} shows the original function for $k=0.75$ a.u., its fit and the fitting errors obtained with the three different optimizations.
By comparing results of fit (A) and fit (B), we observe that using the dimensionless search allows for more flexibility yielding finally a better fit. The results are even better when allowing for a larger search domain as can be seen from the comparison of (B) and (C) curves.

To better understand the way the Gaussian exponents are changed during the optimization, we show  in table \ref{tab:optimized_exponents} their final optimized values for each fit, together with their initial selected values as defined by eq.~\eqref{eq:initial_exp}. This comparison is instructive because it turns out that some large real parts of the exponents remain almost unchanged in fit (A). In the case of fit (B), the research bounds are relative to the initial values; while this leads  to more flexibility, we can still remark that some exponents saturate at, or close to, their minimum allowed bound. Fit (C) is even more flexible with only a few saturated variables.

\begin{table}

\caption{ \label{tab:fitting_parametersABC}
Optimization parameters for three runs of the fitting algorithm. Run A corresponds to an optimization with the same bounds for all the Gaussian exponents. Runs B and C use the dimensionless search as explained in the text. Run C allows the real part of the exponents to go to smaller values. }

{\scriptsize

\begin{tabular}{|p{2.8cm}|p{2.cm}|p{3.5cm}|p{3.7cm}|}

\hline
Fit	&		(A)	&	(B)	&	(C)	\\
\hline
Optimization method & Absolute value & Dimensionless & Dimensionless \\
\hline
\textit{Initial values:} &&& \\
$a=\text{Re}(\alpha_1^{(0)}))$	&	$10^{-4}$	&	$10^{-4}$	&	$10^{-4}$	\\
$b=\text{Re}(\alpha_{30}^{(0)})$		&	$10^{2}$	&	$10^{2}$	&	$10^{2}$	\\
$\text{Im}(\alpha_s^{(0)}), \; \forall s$	&	0	&	$10^{-2}$	&	$10^{-2}$	\\
\hline
\textit{Research bounds:}  & & & \\
 	$\text{Re}(\alpha_s) \in$	&	$[10^{-4} ; 10^{4}], \; \forall s$	&	$[0.1\;  \text{Re}(\alpha_s^{(0)}) ; 10 \; \text{Re}(\alpha_s^{(0)})]$	&	$[10^{-2} \; \text{Re}(\alpha_s^{(0)}) ; 10 \;\text{Re}(\alpha_s^{(0)})]$	\\
		$\text{Im}(\alpha_s) \in$	&	$[-0.1 ; 0.1], \; \forall s$	&	$[-10  \; \text{Im}(\alpha_s^{(0)}) ; 10 \; \text{Im}(\alpha_s^{(0)})]$	&	$[-10  \; \text{Im}(\alpha_s^{(0)}) ; +10 \; \text{Im}(\alpha_s^{(0)})]$	\\
\hline
\textit{Trust region radius:}   & & & \\
		$\Delta_{beg}$	&	$10^{-2}$	&	$10^{-1}$	&	$10^{-1}$	\\
		$\Delta_{end}$	&	$10^{-7}$	&	$10^{-7}$	&	$10^{-7}$	\\
\hline
\textit{Objective function:}	 &&& \\
Total 	&	$7.39 \times 10^{-3} $	&	$4.73 \times 10^{-4} $	&	$1.83 \times 10^{-5}$	\\
w/o penalty term 	&	$7.27 \times 10^{-3} $	&	$4.06 \times 10^{-4} $	&	$ 9.88 \times 10^{-6} $	\\

\hline
\end{tabular}

}

\end{table}

\begin{landscape}

\begin{table}

\caption{ \label{tab:optimized_exponents}
Gaussian exponents for three runs of the fitting algorithm (see optimization parameters in table \ref{tab:fitting_parametersABC}). The second column gives the choice of initial values (real part) before optimization following equation \eqref{eq:initial_exp}. The initial value of $\text{Im}(\alpha_s^{(0)})$ is $0$ for run A and $10^{-2}$ for runs B and C. The following columns give the optimal exponents.
Bold numbers correspond to exponents which remain practically unchanged during the optimization. Underlined numbers are exponents which have reached one of the imposed research bound and saturate close to, or at this boundary value. For run B, several exponents saturate at their minimum allowed value.
}

{\scriptsize

\begin{tabular}{|c|c|c|c|c|c|c|}

\hline
 &	$\text{Re}(\alpha_s^{(0)})$	& $\text{Im}(\alpha_s^{(0)})$& $\text{Im}(\alpha_s^{(0)})$	&	Optimized $\alpha_s$ & Optimized $\alpha_s$ & Optimized  $\alpha_s$  \\
$s$ &	 (A and B)	& (A)		& (B and C)		& (A)		& (B)		&	(C)		\\
	 \hline
  &&&&&& \\
1 & $	1.000\times 10^{-4}	$	&	$	0	$	&	$	10^{-2}	$	&	$	1.000\times 10^{-4}		+1.245\times 10^{-2}\imath	$	&	$	\underline{1.000\times 10^{-5}}	+2.357\times 10^{-2}\imath	$	&	$	2.220\times 10^{-4}		+1.207\times 10^{-2}\imath	$	\\
2 & $	1.610\times 10^{-4}	$	&	$	0	$	&	$	10^{-2}	$	&	$	1.296\times 10^{-4}		-2.204\times 10^{-2}\imath	$	&	$	\underline{1.610\times 10^{-5}}	-9.846\times 10^{-3}\imath	$	&	$	3.193\times 10^{-4}		-5.068\times 10^{-3}\imath	$	\\
3 & $	2.592\times 10^{-4}	$	&	$	0	$	&	$	10^{-2}	$	&	$	2.750\times 10^{-3}		+3.721\times 10^{-2}\imath	$	&	$	\underline{2.592\times 10^{-5}}	+8.330\times 10^{-3}\imath	$	&	$	4.543\times 10^{-4}		-2.296\times 10^{-2}\imath	$	\\
4 & $	4.175\times 10^{-4}	$	&	$	0	$	&	$	10^{-2}	$	&	$	5.705\times 10^{-4}		-1.416\times 10^{-2}\imath	$	&	$	1.698\times 10^{-3}				+3.289\times 10^{-2}\imath	$	&	$	6.384\times 10^{-4}		+2.174\times 10^{-2}\imath	$	\\
5 & $	6.723\times 10^{-4}	$	&	$	0	$	&	$	10^{-2}	$	&	$	9.288\times 10^{-4}		-2.797\times 10^{-2}\imath	$	&	$	2.157\times 10^{-3}				-3.409\times 10^{-2}\imath	$	&	$	8.909\times 10^{-4}		-2.341\times 10^{-2}\imath	$	\\
6 & $	1.082\times 10^{-3}	$	&	$	0	$	&	$	10^{-2}	$	&	$	4.434\times 10^{-4}		+1.585\times 10^{-2}\imath	$	&	$	2.703\times 10^{-3}				-2.323\times 10^{-2}\imath	$	&	$	1.232\times 10^{-3}		-2.379\times 10^{-2}\imath	$	\\
7 & $	1.743\times 10^{-3}	$	&	$	0	$	&	$	10^{-2}	$	&	$	8.981\times 10^{-3}		-4.295\times 10^{-2}\imath	$	&	$	3.393\times 10^{-3}				-2.296\times 10^{-2}\imath	$	&	$	1.691\times 10^{-3}		-2.434\times 10^{-2}\imath	$	\\
8 & $	2.807\times 10^{-3}	$	&	$	0	$	&	$	10^{-2}	$	&	$	1.153\times 10^{-3}		+2.264\times 10^{-2}\imath	$	&	$	4.267\times 10^{-3}				-1.834\times 10^{-2}\imath	$	&	$	2.303\times 10^{-3}		+2.962\times 10^{-2}\imath	$	\\
9 & $	4.520\times 10^{-3}	$	&	$	0	$	&	$	10^{-2}	$	&	$	1.676\times 10^{-4}		-1.317\times 10^{-2}\imath	$	&	$	5.376\times 10^{-3}				+1.588\times 10^{-2}\imath	$	&	$	3.136\times 10^{-3}		-6.163\times 10^{-3}\imath	$	\\
10 & $	7.278\times 10^{-3}	$	&	$	0	$	&	$	10^{-2}	$	&	$	7.420\times 10^{-4}		+2.920\times 10^{-2}\imath	$	&	$	6.810\times 10^{-3}				+4.110\times 10^{-2}\imath	$	&	$	4.245\times 10^{-3}		-3.461\times 10^{-2}\imath	$	\\
11 & $	1.172\times 10^{-2}	$	&	$	0	$	&	$	10^{-2}	$	&	$	3.486\times 10^{-3}		-3.609\times 10^{-2}\imath	$	&	$	8.620\times 10^{-3}				-4.338\times 10^{-2}\imath	$	&	$	5.625\times 10^{-3}		+3.660\times 10^{-2}\imath	$	\\
12 & $	1.887\times 10^{-2}	$	&	$	0	$	&	$	10^{-2}	$	&	$	7.169\times 10^{-3}		+4.615\times 10^{-2}\imath	$	&	$	1.082\times 10^{-2}				+1.200\times 10^{-2}\imath	$	&	$	7.458\times 10^{-3}		+1.664\times 10^{-2}\imath	$	\\
13 & $	3.039\times 10^{-2}	$	&	$	0	$	&	$	10^{-2}	$	&	$	1.370\times 10^{-2}		-4.688\times 10^{-2}\imath	$	&	$	1.370\times 10^{-2}				+1.315\times 10^{-2}\imath	$	&	$	9.862\times 10^{-3}		+1.550\times 10^{-2}\imath	$	\\
14 & $	4.893\times 10^{-2}	$	&	$	0	$	&	$	10^{-2}	$	&	$	1.774\times 10^{-2}		-6.515\times 10^{-2}\imath	$	&	$	1.779\times 10^{-2}				+5.557\times 10^{-2}\imath	$	&	$	1.299\times 10^{-2}		+4.624\times 10^{-2}\imath	$	\\
15 & $	7.880\times 10^{-2}	$	&	$	0	$	&	$	10^{-2}	$	&	$	2.112\times 10^{-2}		+5.848\times 10^{-2}\imath	$	&	$	2.164\times 10^{-2}				+4.104\times 10^{-2}\imath	$	&	$	1.644\times 10^{-2}		-4.075\times 10^{-2}\imath	$	\\
16 & $	1.268\times 10^{-1}	$	&	$	0	$	&	$	10^{-2}	$	&	$	2.555\times 10^{-2}		+6.278\times 10^{-2}\imath	$	&	$	2.632\times 10^{-2}				-4.934\times 10^{-2}\imath	$	&	$	2.085\times 10^{-2}		-3.780\times 10^{-2}\imath	$	\\
17 & $	2.043\times 10^{-1}	$	&	$	0	$	&	$	10^{-2}	$	&	$	4.648\times 10^{-2}		-9.969\times 10^{-2}\imath	$	&	$	3.227\times 10^{-2}				-4.799\times 10^{-2}\imath	$	&	$	2.646\times 10^{-2}		-5.017\times 10^{-2}\imath	$	\\
18 & $	3.290\times 10^{-1}	$	&	$	0	$	&	$	10^{-2}	$	&	$	5.649\times 10^{-2}		+9.799\times 10^{-2}\imath	$	&	$	4.009\times 10^{-2}				-5.691\times 10^{-2}\imath	$	&	$	3.344\times 10^{-2}		+5.658\times 10^{-2}\imath	$	\\
19 & $	5.298\times 10^{-1}	$	&	$	0	$	&	$	10^{-2}	$	&	$	1.658\times 10^{-1}		-1.000\times 10^{-1}\imath	$	&	$	\underline{5.298\times 10^{-2}}	+7.747\times 10^{-2}\imath	$	&	$	4.283\times 10^{-2}		+5.080\times 10^{-2}\imath	$	\\
20 & $	8.531\times 10^{-1}	$	&	$	0	$	&	$	10^{-2}	$	&	$	2.036\times 10^{-1}		+1.000\times 10^{-1}\imath	$	&	$	\underline{8.531\times 10^{-2}}	-7.619\times 10^{-4}\imath	$	&	$	5.548\times 10^{-2}		-6.647\times 10^{-2}\imath	$	\\
21 & $	1.373				$	&	$	0	$	&	$	10^{-2}	$	&	$	1.259					+1.331\times 10^{-2}\imath	$	&	$	\underline{1.373\times 10^{-1}}	-1.294\times 10^{-2}\imath	$	&	$	7.400\times 10^{-2}		+2.421\times 10^{-2}\imath	$	\\
22 & $	2.212				$	&	$	0	$	&	$	10^{-2}	$	&	$	\bf{2.196}				+3.515\times 10^{-3}\imath	$	&	$	\underline{2.212\times 10^{-1}}	-1.545\times 10^{-2}\imath	$	&	$	9.970\times 10^{-2}		-2.150\times 10^{-2}\imath	$	\\
23 & $	3.562				$	&	$	0	$	&	$	10^{-2}	$	&	$	\bf{3.497}				+3.162\times 10^{-2}\imath	$	&	$	\underline{3.562\times 10^{-1}}	-1.946\times 10^{-2}\imath	$	&	$	1.370\times 10^{-1}		+\underline{1.000\times 10^{-1}}\imath	$	\\
24 & $	5.736				$	&	$	0	$	&	$	10^{-2}	$	&	$	\bf{5.712}				+1.232\times 10^{-2}\imath	$	&	$	\underline{5.736\times 10^{-1}}	-1.930\times 10^{-2}\imath	$	&	$	1.899\times 10^{-1}		+5.366\times 10^{-2}\imath	$	\\
25 & $	9.236				$	&	$	0	$	&	$	10^{-2}	$	&	$	\bf{9.247}				+5.392\times 10^{-2}\imath	$	&	$	\underline{9.236\times 10^{-1}}	-1.260\times 10^{-2}\imath	$	&	$	2.669\times 10^{-1}		+3.451\times 10^{-2}\imath	$	\\
26 & $	1.487\times 10^{1}	$	&	$	0	$	&	$	10^{-2}	$	&	$	\bf{1.486\times 10^{1}}	-4.114\times 10^{-2}\imath	$	&	$	\underline{1.487}				+6.921\times 10^{-3}\imath	$	&	$	1.367					-3.962\times 10^{-3}\imath	$	\\
27 & $	2.395\times 10^{1}	$	&	$	0	$	&	$	10^{-2}	$	&	$	\bf{2.391\times 10^{1}}	+4.360\times 10^{-3}\imath	$	&	$	\underline{2.395}				+2.988\times 10^{-2}\imath	$	&	$	4.617					+6.191\times 10^{-3}\imath	$	\\
28 & $	3.856\times 10^{1}	$	&	$	0	$	&	$	10^{-2}	$	&	$	\bf{3.853\times 10^{1}}	-6.297\times 10^{-3}\imath	$	&	$	\underline{3.856}				+4.114\times 10^{-2}\imath	$	&	$	1.826\times 10^{1}		+2.515\times 10^{-3}\imath	$	\\
29 & $	6.210\times 10^{1}	$	&	$	0	$	&	$	10^{-2}	$	&	$	\bf{6.209\times 10^{1}}	-5.881\times 10^{-3}\imath	$	&	$	\underline{6.210}				+1.397\times 10^{-2}\imath	$	&	$	9.431\times 10^{1}		+\underline{9.976\times 10^{-3}}\imath	$	\\
30 & $	1.000\times 10^{2}	$	&	$	0	$	&	$	10^{-2}	$	&	$	\bf{9.995\times 10^{1}}	-4.414\times 10^{-2}\imath	$	&	$	1.182\times 10^{2}				-1.594\times 10^{-3}\imath	$	&	$	\underline{1.000\times 10^{3}} +8.489\times 10^{-3}\imath	$	\\
  &&&&&& \\
\hline

\end{tabular}

}

\end{table}

\end{landscape}

\subsection{The long-range inconsistency outside the fitting box}

\begin{figure}

\begin{center}

\includegraphics[width=0.8\linewidth]{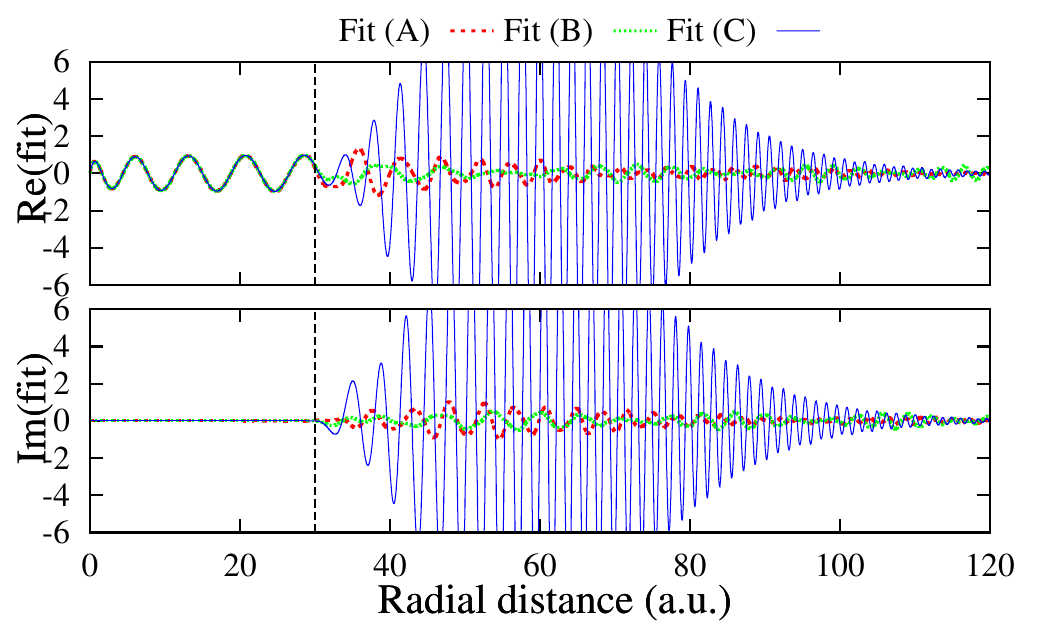}

\end{center}

\caption{ \label{fig2_pb_outside}
Unphysical long-range behaviour of complex Gaussian expansions for Coulomb radial wavefunctions ($k=0.75$ a.u., $l=0$) represented with 30 cGTOs, obtained from three different sets of optimization parameters A, B, C as described in table \ref{tab:fitting_parametersABC}.
The vertical dashed line indicates the radial limit of the fitting box. }

\end{figure}

Let us now have a look at the cGTO representation outside the fitting box. Does the cGTO representation maintain the correct oscillations far from the fitting box boundary, fixed here at the relatively large value of 30 a.u.? Does it go rapidly to zero?
Fig. \ref{fig2_pb_outside} shows that the behaviour of fits (A), (B), (C) are equally and reasonably bad outside the fitting box. There is no strong divergence as it has been already observed with previous aborted attempts to use real GTOs in similar problems \cite{ammar_2020}, but still, there are wrong erratic, unphysical, oscillations.
This may or may not be a serious problem, depending on which application one has in mind; for example,  such limited-range representations will not affect
the calculation of integrals involving overall short range integrands.
We can also note that the most flexible optimization of run (C), the one which gave the best approximation within the fitting box,  also becomes the worst at large distances. This is due to the fact that relaxing the constraint on large exponents leads to the appearance of a larger number of small exponents in the basis, associated to more diffuse oscillations which become spurious outside the fitting box.

\section{A complex Gaussian representation respectful of the oscillatory asymptotic behaviour \label{sec:asymptotics} }

\subsection{Indirect fitting strategy using asymptotic factorization
 \label{ssec:indirect_fit}}


In ref. \cite{fiori_2012}, Fiori and Miraglia factorized the plane wave factor   from the three-dimensional Coulomb continuum wavefunction (thus extracting a large part of the oscillations) and then expanded in partial waves the remaining distortion factor, before performing a real GTO representation.
Here, we suggest to take into account more rigorously the whole asymptotic behaviour of the radial function in a new choice of factorization designed to facilitate the cGTO representation while allowing for an easy reconstruction of the exact asymptotic oscillations.
For a general radial function $u_{l,k}(r)$,
solution of eq.~\eqref{eq:fct_disto_DE} for a given central distortion potential $V(r)$, with given momentum $k$ and orbital quantum number $l$,
the asymptotic behaviour features a given phase shift $\delta_l$.
Although the following proposal is general, we shall proceed hereafter with the pure Coulomb case, for which asymptotically the radial function behaves as:
\begin{equation}
\begin{aligned}
	u_{l,k}(r)
	\sim &
	 \sin\left[k r -\eta\ln(2kr)-l\frac{\pi}{2}+\delta_l \right]	
\\
	\sim &
	 \frac{1}{2\imath} \left[e^{\imath[k r -\eta\ln(2kr)-l\frac{\pi}{2}+\delta_l ]}-e^{-\imath[k r -\eta\ln(2kr)-l\frac{\pi}{2}+\delta_l ]}\right],	\end{aligned}	\label{eq:RegCoulFun2}
\end{equation}
where the Coulomb phase shift is $\delta_l = \arg \left( \Gamma(l+1 + \imath \eta) \right)$.

An intuitive,
 simple
proposal consists in trying to extract from the Coulomb radial function the dominating Coulomb asymptotic oscillation (including the logarithmic term). We thus define some factorized function $v^{(1)}_{l,k}(r)$  such that
\begin{equation}
\begin{aligned}
	u_{l,k}(r) & = \sin\left[k r -\eta\ln(2kr)-l\frac{\pi}{2}+\delta_l \right] \; v^{(1)}_{l,k}(r) \\
	&  \approx \sin\left[k r -\eta\ln(2kr)-l\frac{\pi}{2}+\delta_l \right]	 \;
\sum_{s=1}^{N}
		\left[ c_s ^{(1)} \right]_{l,k}  e^{-\left[ \alpha_s^{(1)} \right]_{l} \, r^2},
	\label{eq:cG_def_modified1}
\end{aligned}
\end{equation}
the second line indicating that it is  the transformed function
\begin{equation}
v^{(1)}_{l,k}(r) \equiv u_{l,k}(r) / \sin \left[k r -\eta\ln(2kr)-l\frac{\pi}{2}+\delta_l \right],
\end{equation}
instead of $u_{l,k}(r)$, that we are representing with cGTOs.
However their is a serious problem with this first suggestion since the $v^{(1)}_{l,k}$ function presents obvious divergences at values of $r$ cancelling the sine  factor. In no way a cGTO expansion could be able to mimic such divergences.

We can easily circumvent this problem by translating the original oscillating function $u_{l,k}(r)$ by some offset parameter $C>1$ before factorizing. We thus consider the function  $ u_{l,k}^{\text{offset}} $ such that
\begin{equation}
\begin{aligned}
	u_{l,k}^{\text{offset}} (r) & \equiv  C + u_{l,k}(r) \\
 & = \left\{ C + \sin\left[k r -\eta\ln(2kr)-l\frac{\pi}{2}+\delta_l \right] \right\}
  \; v^{(2)}_{l,k}(r).
	\label{eq:cG_def_modified2}
\end{aligned}
\end{equation}
The translated function $u_{l,k}^{\text{offset}}$ now oscillates asymptotically around the offset value $C$, while the factorized function
$v^{(2)}_{l,k}(r)$ oscillates around $1$ without singularities. To obtain a distortion function oscillating around zero instead, the best choice we propose is to now vertically translate the $v_{l,k}^{(2)}$ function and define:
\begin{equation}
\begin{aligned}
\tilde{u}_{l,k}(r) & \equiv  v^{(2)}_{l,k}(r)  - 1 \\
& =  \frac{ C + u_{l,k}(r) }{C + \sin \left[k r -\eta\ln(2kr)-l\frac{\pi}{2}+\delta_l \right] }  - 1.
\end{aligned}
\label{eq:distortion_factor}
\end{equation}
This choice of distortion function is the most appropriate candidate for cGTO representation because $\tilde{u}_{l,k}(r)$ tends to zero and oscillates around zero when the original function $u_{l,k}(r)$ gradually reaches its sine asymptotic regime.

The proposed strategy runs as follows. For a given function $u_{l,k}(r)$, we use eq. \eqref{eq:distortion_factor} to calculate the distortion function $\tilde{u}_{l,k}$ and then fit it by a cGTO expansion,
\begin{equation}
\tilde{u}_{l,k}(r) \approx \sum_{s=1}^{N}
		\left[ \tilde{c}_s  \right]_{l,k}  e^{-\left[ \tilde{\alpha}_s \right]_{l} \, r^2}.
\label{eq:fit_distortion_factor}
\end{equation}
Finally, the ``original" wavefunction $u_{l,k}$ is reconstructed from the distortion function $\tilde{u}_{l,k}$:
\begin{equation}
\begin{aligned}
u_{l,k} (r)
\approx  &
 \left\{ C + \sin \left[k r -\eta\ln(2kr)-l\frac{\pi}{2}+\delta_l \right]  \right\}
\sum_{s=1}^{N}
		\left[ \tilde{c}_s  \right]_{l,k}  e^{-\left[ \tilde{\alpha}_s \right]_{l} \, r^2} \\
& + \sin \left[k r -\eta\ln(2kr)-l\frac{\pi}{2}+\delta_l \right].
\end{aligned}
\label{eq:reconstruction}
\end{equation}
As will be shown in section \ref{sec:integrals}, this format does not generate any complications in the calculation of typical integrals that appear, e.g., in the context of electron detachment processes.

\begin{figure}

\begin{center}

\includegraphics[width=0.8\linewidth]{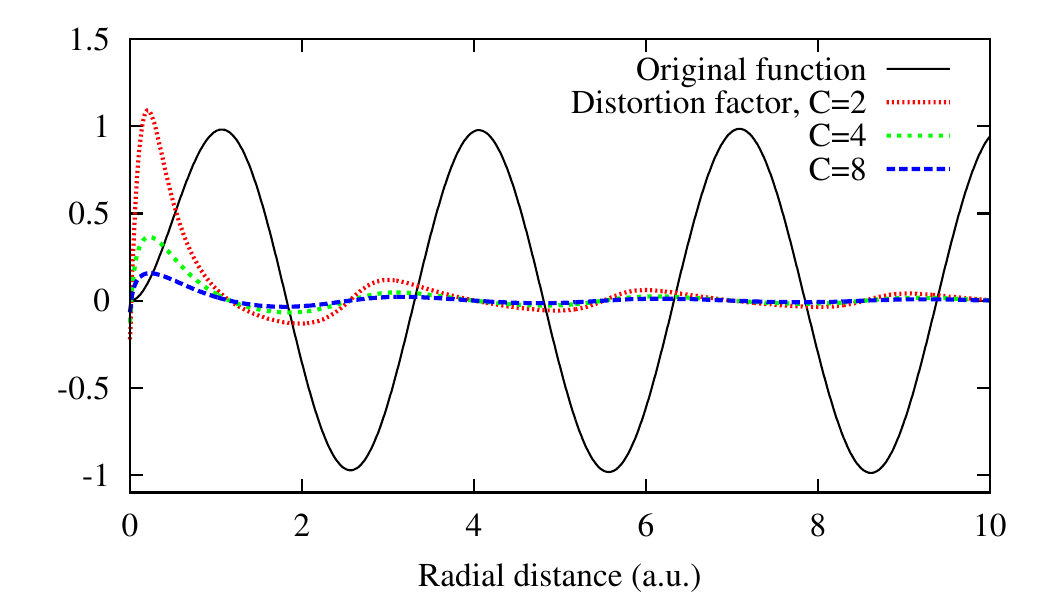}

\end{center}

\caption{ \label{fig:distortion_factors}
Original Coulomb radial function $u_{l,k}(r)$ for $k=2$ a.u. and $l=1$ (black solid line),
together with their associated asymptotic distortion factor $\tilde{u}_{l,k}(r)$ as defined in eq.~\eqref{eq:distortion_factor}, for three selected values of the offset parameter ($C=2$, red dotted line; $C=4$, green dashed line; $C=8$, blue long-dashed line).
 }

\end{figure}


Let us now illustrate these distortion functions $\tilde{u}_{l,k}(r)$ in the case of radial Coulomb functions.
Fig. \ref{fig:distortion_factors} shows that they are clearly localized and oscillating around zero as expected, contrary to the original function which oscillates until infinity. This was precisely the aim of the proposed transformation.
Fig. \ref{fig:distortion_factors} also illustrates the fact that the absolute value of $\tilde{u}_{l,k}(r) $ decreases with increasing offset parameter $C$, leading to smoother variations.

\begin{table}

\caption{Optimization parameters and convergence of the objective function using the indirect fitting method with distortion functions $\tilde{u}_{l=1,k}(r)$ defined in eq.~\eqref{eq:distortion_factor}.
\label{tab:par_conv_distortion_function}
}

{\scriptsize

\begin{tabular}{|p{2.8cm}|p{1.3cm} p{1.3cm} p{1.3cm}|p{1.3cm} p{1.3cm} p{1.3cm}|}
\hline
Momentum $k$ (a.u.) & 1  & 1 & 1  &  2 & 2  & 2  \\
\hline
Offset $C$  &		2	&	4	&	8  & 2	&	4	&	8 \\
\hline
Optimization method & Dimensionless & & & Dimensionless & & \\
\hline
\textit{Initial values:} &&& &&&\\
$a=\text{Re}(\alpha_1^{(0)}))$	&	$3 \times 10^{-3}$	&&& $3 \times 10^{-3}$ &&\\
$b=\text{Re}(\alpha_{30}^{(0)})$		&	$5 \times 10^{2}$	&&& $5 \times 10^{2}$ &&  \\
$\text{Im}(\alpha_s^{(0)}), \; \forall s$	&	$10^{-2}$	&&& $10^{-2}$ &&  \\
\hline
\textit{Research bounds:}  & & & & & & \\
 	$\text{Re}(\alpha_s) \in$	&	$[0.1\;  \text{Re}(\alpha_s^{(0)}) ; 10 \; \text{Re}(\alpha_s^{(0)})]$	& & &	$[0.1\;  \text{Re}(\alpha_s^{(0)}) ; 10 \; \text{Re}(\alpha_s^{(0)})]$ &	&\\
		$\text{Im}(\alpha_s) \in$	& $[-20  \; \text{Im}(\alpha_s^{(0)}) ; 20 \; \text{Im}(\alpha_s^{(0)})]$	&&&	$[-20  \; \text{Im}(\alpha_s^{(0)}) ; 20 \; \text{Im}(\alpha_s^{(0)})]$	&& \\
\hline
\textit{Trust region radius:}   & & &&&&\\
		$\Delta_{beg}$	&	$10^{-1}$	&	&	& $10^{-1}$ && \\
		$\Delta_{end}$	&	$10^{-7}$	&	&	& $10^{-7}$ && \\
\hline
\textit{Objective function:}	 &&& &&& \\
Total 	&	$5.5 \times 10^{-4} $	&	$2.0 \times 10^{-4} $	&	$1.1 \times 10^{-5}$	& $1.6 \times 10^{-3} $ & $5.2 \times 10^{-4} $ & $2.9 \times 10^{-4} $\\
w/o penalty term 	&	$5.4 \times 10^{-4} $	&	$1.9 \times 10^{-4} $	&	$ 1.0 \times 10^{-5} $	& $1.6 \times 10^{-3} $ & $4.9 \times 10^{-4} $ & $2.8 \times 10^{-4} $ \\
\hline
\end{tabular}
}

\end{table}

\begin{figure}

\begin{center}

\includegraphics[width=0.7\linewidth]{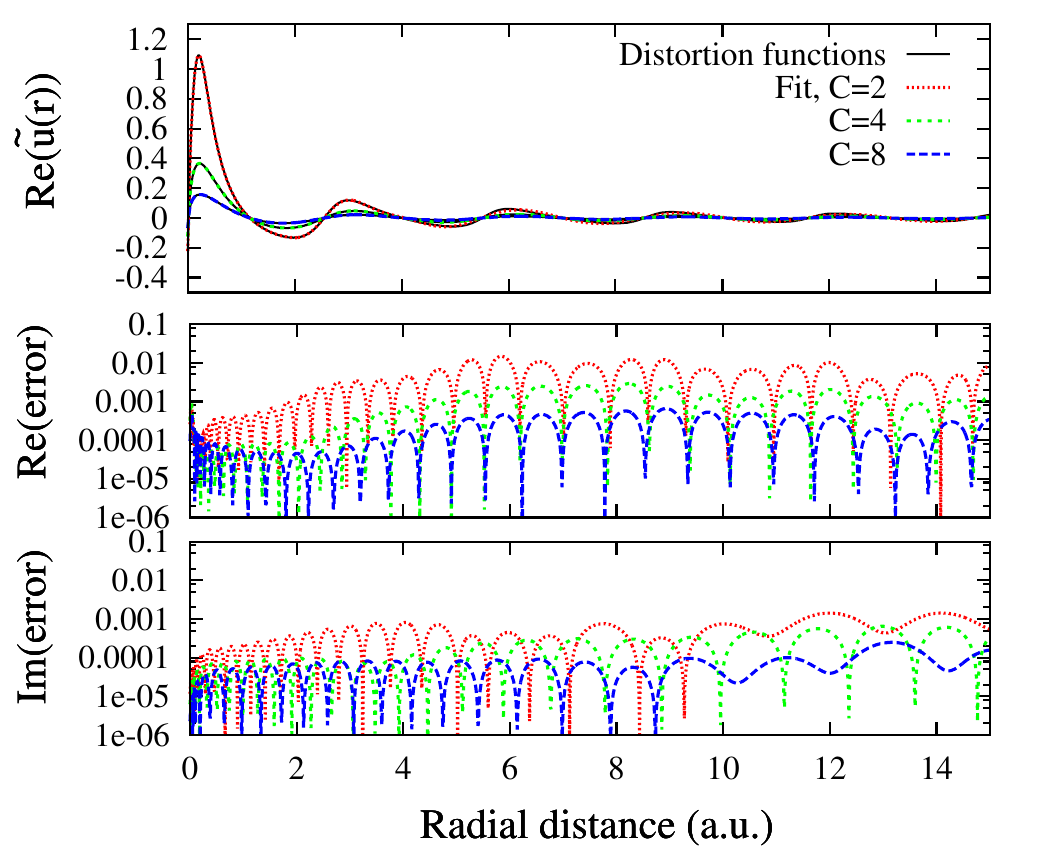}

\end{center}

\caption{ \label{fig:fit_transformed}
Distortion functions $\tilde{u}_{l,k}(r)$ (for $k=2$ a.u., $l=1$), together with their complex Gaussian representations using 30 cGTOs, obtained with three different values of the offset constant $C$.
The upper panel shows the function and its different Gaussian fits, the second and third panels show the real and imaginary parts of the error, respectively.}

\end{figure}

\begin{figure}

\begin{center}

\includegraphics[width=0.7\linewidth]{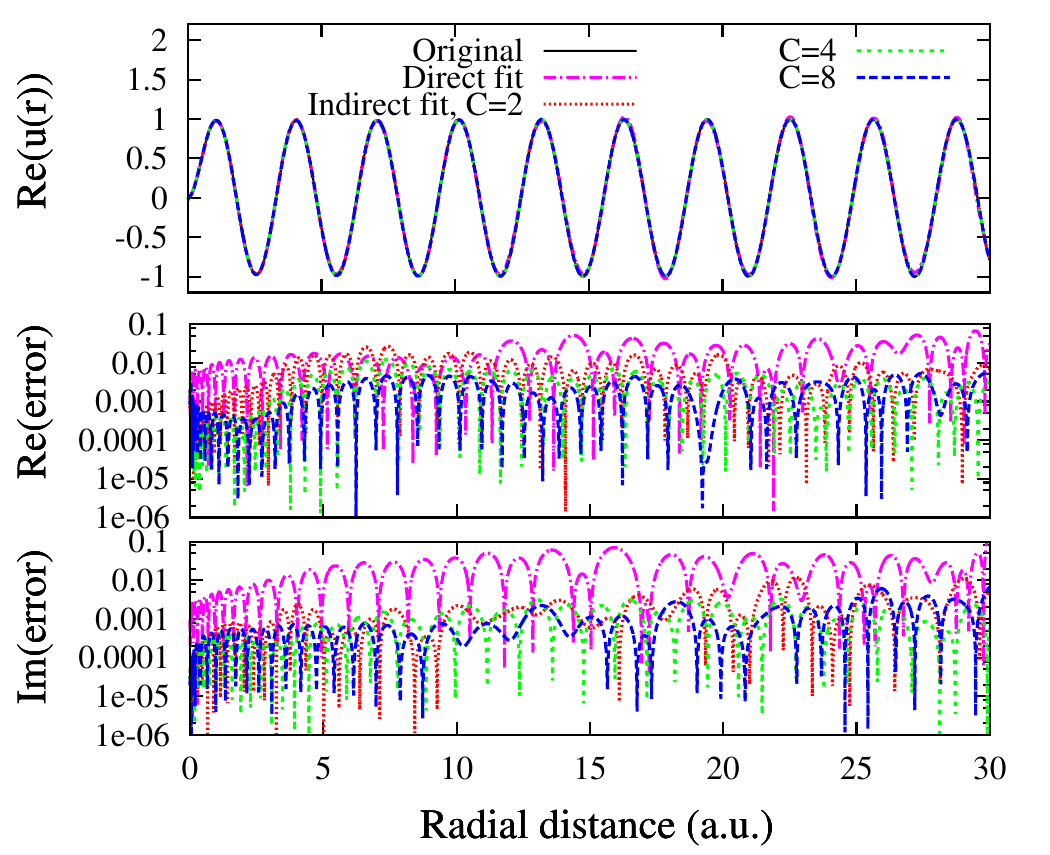}

\end{center}

\caption{ \label{fig:reconstruction}
Reconstruction of the original radial continuum wavefunction $u_{l,k}(r)$ ($k=2$ a.u., $l=1$) within the fitting box, using eq.~\eqref{eq:reconstruction} with 30 cGTOs. We show the results obtained with three different values of the offset constant $C$ (red dotted line for $C=2$, green dashed line for $C=4$, blue long-dashed line for $C=8$. See table \ref{tab:par_conv_distortion_function} for the details). The indirect fit is also compared to an example of direct fit (magenta dotted-dashed line).
The upper panel shows the function and its different Gaussian fits, the second and third panels show the real and imaginary parts of the error, respectively.}

\end{figure}

Each function has been fitted individually with 30 complex Gaussian basis functions
to ensure consistent comparisons with direct fitting results.
Table \ref{tab:par_conv_distortion_function} summarizes the numerical parameters used to optimize the objective function for $l=1$ and $k=1$ or 2 a.u., and shows the convergence for three $C$ values.
We obtain overall very good quality fits, as shown in Fig. \ref{fig:fit_transformed} for $k=2$ a.u..
The fitting error clearly decreases with increasing offset constant $C$. However, we must remember that we are not working on the ``true" function so that the significance of this improvement is not obvious because the absolute value of the distortion functions also decreases with increasing $C$. When $C$ is larger, the distortion function $\tilde{u}_{l,k}(r)$ is to be multiplied afterwards by larger numbers to reconstruct the original function, so the observed advantage could be jeopardized by larger error propagation.
We show in Fig.~\ref{fig:reconstruction} an example of Coulomb function reconstructed with the indirect fitting strategy:
the errors are smaller and still decrease a little with increasing offset constants $C$.

\begin{figure}

\begin{center}

\includegraphics[width=0.75\linewidth]{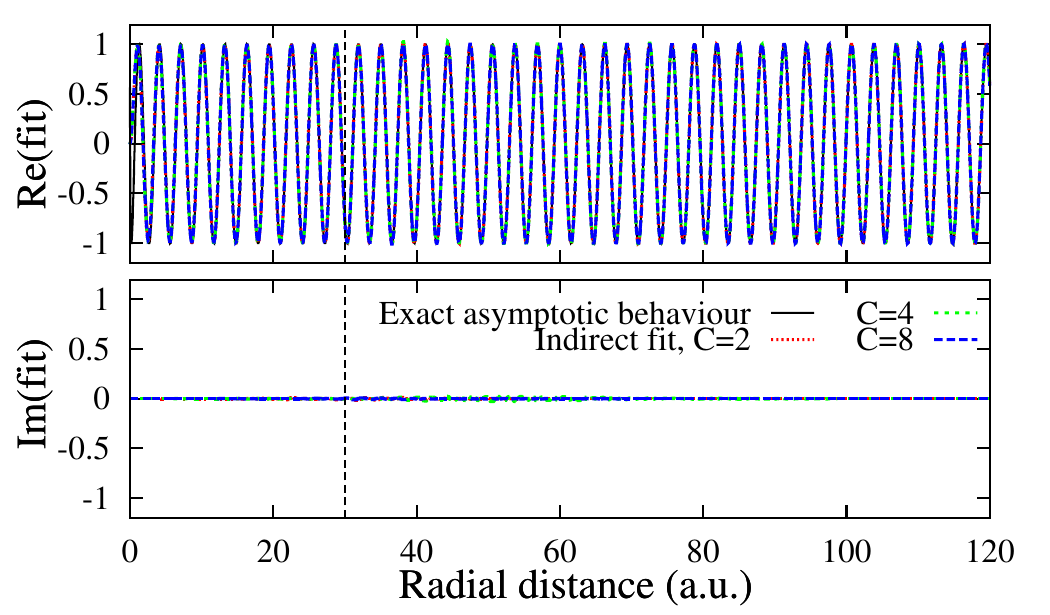}

\end{center}

\caption{ \label{fig:asymptotics}
Asymptotic behaviour of the reconstructed wavefunction ($k=2$ a.u., $l=1$) using the indirect fitting method described in subsection \ref{ssec:indirect_fit} and comparison to the exact theoretical asymptotic oscillations (shown in black continuous line).
We use eq.~\eqref{eq:reconstruction} with cGTOs representation of distortion functions $\tilde{u}_{l,k}(r)$, obtained with three different values of the offset constant $C$ (red dotted line for $C=2$, green dashed line for $C=4$, blue long-dashed line for $C=8$. See table \ref{tab:par_conv_distortion_function} for the details). }

\end{figure}

Most importantly, we have to check the asymptotic behaviour of the reconstructed functions outside the fitting box.
Fig. \ref{fig:asymptotics} shows that, for the three offset constant values, the oscillatory asymptotic behaviour is very well recovered by the indirect fitting method. This is what we expected from the proposed transformation and is in contrast to the unphysical behaviours observed in Fig. \ref{fig2_pb_outside}.
While the figure shows the correct oscillations up to 120 a.u., they are actually recovered up to infinity, by construction.
We have thus successfully obtained a complex Gaussian representation of a continuum radial function respectful of the correct asymptotic oscillations of the original function.
We also would like to emphasize the fact that the proposed method can be applied with no additional difficulty to radial functions associated with more general potentials (\textit{e.g.} Coulombic potential with screening effects, distorted potentials) as long as the model wavefunction is numerically known on a grid together with the associated phase shift (produced,  for example, by the Radial package \cite{salvat2019}).

\subsection{Strategy for the choice of the offset constant $C$}


We now look at how to choose an appropriate value of the offset constant $C$. 
It is clear that $C$ should be chosen large enough to dampen the distortion function $\tilde{u}_{l,k}(r)$ so that the reconstructed physical function recovers the exact asymptotic behaviour up to infinity (last term of eq. \eqref{eq:reconstruction}).  Equivalently, this implies that the fit of eq. \eqref{eq:fit_distortion_factor} correctly goes to zero outside the fitting box.
At the same time, the offset constant $C$ should not be too large to avoid an excessive drop to zero within the fitting box of the distortion factor; this could  potentially lead to strong numerical error amplification in the reconstructed function through the first term of eq.~\eqref{eq:reconstruction}.

\begin{figure}

\begin{center}

\includegraphics[width=0.8\linewidth]{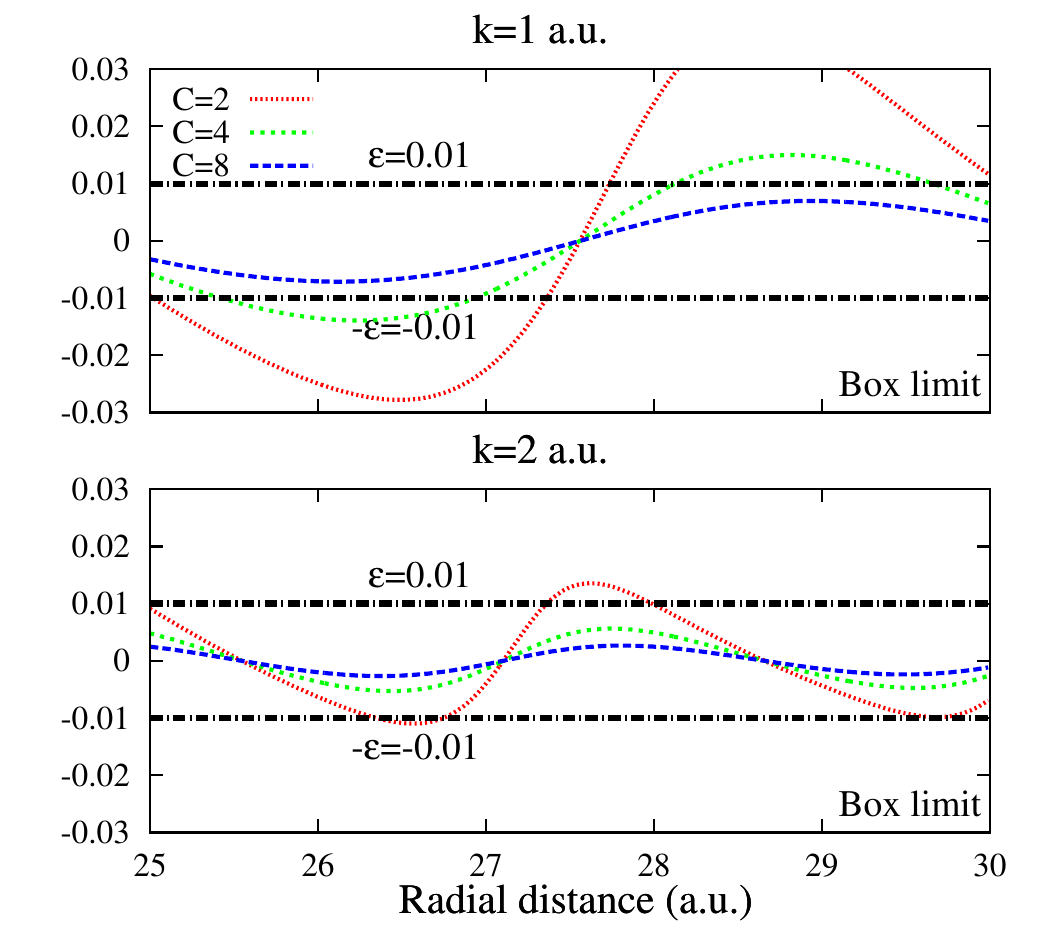}

\end{center}

\caption{ \label{fig:zoom_distortion_factors}
Zooms on the distortion factor $\tilde{u}_{l,k}(r)$ of eq.~\eqref{eq:distortion_factor} close to the fitting box limit $r_{\text{box}}=30$ a.u., for $k=1$ and $k=2$ as indicated, $l=1$ and three selected values of the offset parameter ($C=2$, red dotted line; $C=4$, green dashed line; $C=8$, blue long-dashed line).
The horizontal lines show the convergence criterion fixed here at $\varepsilon = 10^{-2}$.
 }

\end{figure}

Here we propose a practical and systematic way of estimating the order of magnitude for the 'minimal' $C$ value for a given energy $E=k^2/2$ and for a selected size of the fitting box ($r_{\text{box}}$). The criterion we take to guide us is to ask for the distortion function $\tilde{u}_{l,k}(r)$ to remain smaller (in absolute value) than some given tolerance threshold $\varepsilon$ outside the fitting box, i.e. for $r > r_{\text{box}}$; this, simultaneously, ensures a good connection between the fitting box and the asymptotic domain.
Fig. \ref{fig:zoom_distortion_factors} illustrates the behaviour of the distortion function $\tilde{u}_{l,k}(r)$
 with $r_{\text{box}} = 30$ a.u., $k=1$ or $2$ a.u. and varying $C$ values. We can see that the decreasing shape of the envelope  depends on the combined values of $k$ and $C$, and remains within the tolerance $\varepsilon$, here chosen as $10^{-2}$, only for a sufficiently large $C$ value.

Hereafter, the rule of thumb for choosing the offset constant $C$ is based on the particular case of a Coulomb wave.
Including first order corrective terms in $1/r$ (see, e.g., section 14.5 of \cite{abramowitz_1964}),  
the Coulomb radial wavefunction behaves asymptotically as
\begin{equation}
	u_{l,k}(r)
	\approx
	 f  \sin\left[ \theta_{l,k}(r) \right] 	+
	 g  \cos\left[ \theta_{l,k}(r) \right]
 \label{eq:asympt_detailed}
\end{equation}
with
\begin{equation}
\begin{aligned}
f & = 1 + \frac{\eta}{2 k} \frac{1}{r}, \\
g & = \frac{ l (l+1) + \eta^2 }{2k} \frac{1}{r}, \\
\theta_{l,k}(r) & = k r -\eta\ln(2kr)-l\frac{\pi}{2}+\delta_l .
\end{aligned}
\end{equation}
Inserting \eqref{eq:asympt_detailed} in eq. \eqref{eq:distortion_factor}, we get, for large values of $r$,
\begin{equation}
\begin{aligned}
\tilde{u}_{l,k}(r)
& \approx
\left(
\frac{ \eta \sin\left[ \theta_{l,k}(r) \right]
+\left(l (l+1) + \eta^2\right) 
\cos\left[ \theta_{l,k}(r) \right]
}{
C + \sin\left[ \theta_{l,k}(r) \right]
}
\right)
\;
\frac{1}{2kr}.
\end{aligned}	\label{eq:asympt_detailed_2}
\end{equation}
We can write an upper bound for each term of the oscillating factor.
When the cosine is 0 and the sine is -1, we get $\frac{\eta}{2 kr (C -1)} $ from the first term,
and when the cosine is 1 and the sine is 0, we get $\frac{ l (l+1) + \eta^2 }{2kr C} $ from the second term.
We impose, at the fitting box limit, an upper bound value $\varepsilon$ for each subcase, \textit{i.e.}
\begin{equation}
\frac{\vert \eta \vert}{2 \, k \, r_{\text{box}} (C-1)} < \varepsilon
\end{equation}
and
\begin{equation}
\frac{ l (l+1) + \eta^2 }{2 \, k \, r_{\text{box}} \, C} < \varepsilon.
\end{equation}
This gives two complementary estimates for minimum values of $C$ which should be chosen according to one of the following criteria,
\begin{equation}
C  \gtrsim  \frac{z}{4\, E \, r_{\text{box}} \, \varepsilon} + 1
\end{equation}
or
\begin{equation}
C   \gtrsim \frac{ l (l+1) + \eta^2 }{2 \, \sqrt{2E} \, r_{\text{box}} \, \varepsilon} .
\end{equation}
Depending on the values of $k$ and $l$, one criterion may be more restrictive than the other.
As explained above, taking a $C$ value much bigger than this estimate brings no major benefit, and it may even lead to error amplification in reconstructing the original wavefunction.

As an illustration of this rule of thumb, consider the numerical values corresponding to the cases of Fig. \ref{fig:zoom_distortion_factors}. For $k=1$ a.u., $r_{\text{box}}=30$ a.u., $\varepsilon = 10^{-2}$, we get either $C \gtrsim 3$ or $C \gtrsim 5$ ; for $k=2$ a.u., we obtain either $C \gtrsim 2$ or $C \gtrsim 3$.
This is consistent with what can be observed in Fig.~\ref{fig:zoom_distortion_factors}.


\subsection{Consequences on the calculation of typical state to state integrals \label{sec:integrals}}


The cGTO expansions under consideration can be used, for example, in ionization cross-section calculations. The latter require evaluating transition integrals that typically involve some transition operator, a given initial bound electronic state usually represented by Slater-type (STOs) or real GTOs, and a continuum state.
As shown in \cite{ammar_2023,ammar_2024}, representing the continuum radial function by cGTO expansions leads to very practical closed form expressions for all such typical integrals. We show hereafter that this crucial advantage is maintained when using the indirect cGTO representation introduced in section  \ref{ssec:indirect_fit}.

Let us consider, as a case of study, typical radial integrals arising in the case of photoionization cross-section calculations.
Depending of the selected basis functions for the initial state (STOs with exponent $\zeta$ or GTOs with exponent $\gamma$, respectively), the radial integrals take one of the following forms:
\begin{equation}
	{\mathcal{ I}}_{l,k}\left( \zeta, n \right) =
	\int_0^{\infty} \left(u_{l,k}(r)\right)^*  \, r^{n} e^{-\zeta r} \,  d {r}
	\text{,}
	\label{eq:RadInteg_Idef_q0}
\end{equation}
\begin{equation}
	{\mathcal{ J}}_{l,k}\left( \gamma, n \right) =
	\int_0^{\infty} \left(u_{l,k}(r)\right)^*  \, r^{n} e^{-\gamma r^2}  d {r}
	\text{.}
	\label{eq:RadInteg_Jdef_q0}
\end{equation}
For simplicity, in what follows, we will limit our discussion to integral \eqref{eq:RadInteg_Idef_q0} but a similar reasoning can be applied to integral \eqref{eq:RadInteg_Jdef_q0}.

If the continuum state $u_{l,k}(r)$ is replaced by its direct cGTO expansion as defined in eq.~\eqref{eq:cG_def}, $\mathcal{I}_{l,k}$ can be calculated with the following formula~\cite{ammar_2024}:
\begin{equation}
\begin{aligned}
\mathcal{I}_{l,k} \left( \zeta, n \right)
& \approx \sum_{s=1}^{N} \left[ c_s \right]_{l,k}^*
	 \int_0^{\infty}  e^{-\left[ \alpha_s \right]_{l}^* \, r^2} e^{-\zeta r} r^{n} d {r}	\\
& = \sum_{s=1}^{N} \left[ c_s \right]_{l,k}^*
	\mathcal{G} \left( \left[ \alpha_s \right]_l^*, \zeta  , n\right)
	\text{,}
\end{aligned}
		\label{eq:RadInteg_Jgauss}
\end{equation}
where the $\mathcal{G}$ integrals are explicitly calculated using one of the two expressions
\begin{equation}
\begin{aligned}
	\mathcal{G} \left( \alpha , \zeta, n \right)
		&= \frac{ \Gamma\left(n+1\right) }{(2 \alpha)^{\frac{n+1}{2}}} \, \,
                        e^{\frac{\zeta^2}{8 \alpha } } \,
			D_{-n-1}\left( \frac{\zeta}{\sqrt{2 \alpha}} \right) \\
		&= \frac{ \Gamma\left(n+1\right) }{(4 \alpha)^{\frac{n+1}{2}}} \, \,
			U\left( \frac{n+1}{2} , \frac{1}{2} ; \frac{\zeta^2}{4 \alpha } \right)
	\text{.}
	\label{eq:Gauss_U}
\end{aligned}
\end{equation}
In eq.~\eqref{eq:Gauss_U},  $D_{\nu}$ is the parabolic cylinder function and $U$ stands for the Tricomi function \cite{bateman1953,gradshteyn2014}. These expressions are also valid for complex parameters $\zeta$ and $n$, with the condition $\text{Re}\left( \frac{\zeta}{\sqrt{2 \alpha}} \right) >0$. Both can be evaluated using standard mathematical packages \cite{johansson2013,nardin1992}.

If we consider instead the proposed Gaussian expansion~\eqref{eq:reconstruction}, consistent with correct asymptotic oscillations,
the calculation is modified as follows.
By substituting \eqref{eq:reconstruction} in eq.~\eqref{eq:RadInteg_Idef_q0}, we obtain
\begin{equation}
\begin{aligned}
& \mathcal{I}_{l,k} \left( \zeta, n \right)
 \approx  \;
 \sum_{s=1}^{N} \left[ \tilde{c}_s  \right]_{l,k}^*
  \left\{
C
\int_0^{\infty}
		  e^{-\left[ \tilde{\alpha}_s \right]_{l}^* \, r^2}
 \, r^{n} e^{-\zeta r} \,  d {r}    \right. \\
& +
\left.
\int_0^{\infty}
e^{-\left[ \tilde{\alpha}_s \right]_{l}^* \, r^2}
\frac{1}{2\imath} \left[e^{\imath[k r - \eta\ln(2kr)-l\frac{\pi}{2}+\delta_l ]}
-e^{-\imath[k r -\eta\ln(2kr)-l\frac{\pi}{2}+\delta_l ]} \right]
 \, r^{n} e^{-\zeta r} \,  d {r}    \right\}\\
& +
\int_0^{\infty}
\frac{1}{2\imath} \left[e^{\imath[k r -\eta\ln(2kr)-l\frac{\pi}{2}+\delta_l ]}-e^{-\imath[k r -\eta\ln(2kr)-l\frac{\pi}{2}+\delta_l ]}\right]
 \, r^{n} e^{-\zeta r} \,  d {r}.  \\
 \end{aligned}
		\label{eq:RadInteg_subst}
\end{equation}
The first term in the braces of \eqref{eq:RadInteg_subst} is simply given by eq.~\eqref{eq:Gauss_U},
\begin{equation}
C
\int_0^{\infty}
		  e^{-\left[ \tilde{\alpha}_s \right]_{l}^* \, r^2}
 \, r^{n} e^{-\zeta r} \,  d {r}
 =
 C \; \mathcal{G} \left( \left[ \tilde{\alpha}_s \right]_{l}^* , \zeta, n \right).
\label{eq:RadInteg_1st_term}
\end{equation}
For the second term in the braces, we remark that $e^{ \pm \imath \eta \ln(2kr) } = e^{ \pm \imath \eta \ln(2k)} \; r^{\imath \eta} $
 so that the two terms in the second integral in eq.~\eqref{eq:RadInteg_subst} can be expressed as
\begin{equation}
\begin{aligned}
&  \pm
\int_0^{\infty}
e^{-\left[ \tilde{\alpha}_s \right]_{l}^* \, r^2}
\frac{1}{2\imath} e^{ \pm \imath[k r - \eta\ln(2kr)-l\frac{\pi}{2}+\delta_l ]}
 \, r^{n} e^{-\zeta r} \,  d {r}   \\
& =
\pm
\frac{1}{2\imath}
e^{\pm \imath \tau_l }
\int_0^{\infty}
e^{-\left[ \tilde{\alpha}_s \right]_{l}^* \, r^2}
e^{-(\zeta \mp \imath k) r} r^{n \mp \imath\eta} d {r} \\
& =
\pm
\frac{1}{2\imath}
e^{\pm \imath \tau_l } \;
\mathcal{G}
\left( \left[ \tilde{\alpha}_s \right]_{l}^* , \zeta \mp \imath k, n \mp \imath\eta \right),
 \end{aligned}
		\label{eq:RadInteg_2d_term}
\end{equation}
where the phase  $\tau_l = -l\frac{\pi}{2}+\delta_l-\eta\ln(2k)$.
The last integral of eq.~\eqref{eq:RadInteg_subst} can be dealt with using the  Gamma function property
$\int_0^{\infty} e^{-xt} t^{z-1} dt = \Gamma(z) / x^z  $
also valid for complex $x$ \cite{gradshteyn2014}, leading to two terms
\begin{equation}
\begin{aligned}
  \pm \int_0^{\infty}
\frac{1}{2\imath}
\left[ e^{ \pm \imath \left[k r -\eta\ln(2kr)-l\frac{\pi}{2}+\delta_l \right] }
 \, r^{n} e^{-\zeta r} \, \right] d {r}
& =
\pm \frac{1}{2\imath}  e^{\pm \imath \tau_l }
\frac{ \Gamma( n +1 \mp \imath \eta  ) }{ ( \zeta  \mp \imath k)^{ n+1 \mp \imath \eta } }.
 \end{aligned}
		\label{eq:RadInteg_3d_term}
\end{equation}
The final formula for integral \eqref{eq:RadInteg_Idef_q0} reads
\begin{equation}
\begin{aligned}
 \mathcal{I}_{l,k} \left( \zeta, n \right)
& \approx  \;
 \sum_{s=1}^{N} \left[ \tilde{c}_s  \right]_{l,k}^*
  \Bigl\{
C \; \mathcal{G} \left( \left[ \tilde{\alpha}_s \right]_{l}^* , \zeta, n \right)
 \\
& +
\frac{1}{2\imath}
\left[
e^{ \imath \tau_l } \;
\mathcal{G}
\left( \left[ \tilde{\alpha}_s \right]_{l}^* , \zeta - \imath k, n - \imath\eta \right)
-
e^{- \imath \tau_l } \;
\mathcal{G}
\left( \left[ \tilde{\alpha}_s \right]_{l}^* , \zeta + \imath k, n + \imath\eta \right)
\right]
 \Bigr\}
\\
& +
\frac{1}{2\imath}
\left[
 e^{ \imath \tau_l }
\frac{ \Gamma( n +1 - \imath \eta  ) }{ ( \zeta  - \imath k)^{ n+1 - \imath \eta } }
-
 e^{- \imath \tau_l }
\frac{ \Gamma( n +1 + \imath \eta  ) }{ ( \zeta  + \imath k)^{ n+1 + \imath \eta } }
\right].
 \end{aligned}
		\label{eq:integral_final}
\end{equation}
Although a little bit heavier, this remains a manageable and easily computable closed form expression.

Should the initial state be described by GTO rather than STO, the final formulation for the integral  \eqref{eq:RadInteg_Jdef_q0} is the same, but $\zeta$ is set to zero, and $ \left[ \tilde{\alpha}_s \right]_{l}^*$ is replaced by $ \left[ \tilde{\alpha}_s \right]_{l}^*+\gamma$.


\begin{table}
\caption{
Numerical values of the integral defined in eq.~\eqref{eq:RadInteg_Idef_q0}. Comparison between a reference numerical calculation using the exact Coulomb function, an analytical calculation using a direct cGTO approximation within a 30 a.u. fitting box (eq.~\eqref{eq:RadInteg_Jgauss}) and an analytical calculation using an indirect cGTO representation respectful of the asymptotic behaviour (eq.~\eqref{eq:integral_final}) with the offset constant  $C=8$.
The selected parameters are $k=1$ a.u., $l=1$, $n=1$ and three different values of the Slater exponent $\zeta$ are compared.
\label{tab:integral_comparison}
}

{\scriptsize

\begin{tabular}{|p{3.cm}|p{0.4cm}|p{5cm} | p{3.5cm} |}
\hline
Method & $\zeta$ & Integral	& Absolute error	\\
\hline
Numerical (reference) 	& 1 & $3.6880147\times 10^{-1} + 0\imath$ 	& -  \\
Analytical,
direct fit 	& 	& $3.6880139\times 10^{-1} + 1.81105182\times 10^{-7}\imath$ 	& $8.5\times 10^{-8} + 1.8\times 10^{-7}\imath$\\
Analytical,
indirect fit & 	& $3.6879384\times 10^{-1} + 2.9282974\times 10^{-9}\imath$	& $7.6\times 10^{-6} + 2.9\times 10^{-9}\imath$  \\
\hline
Numerical (reference) 	& 0.5 & $4.9609361\times 10^{-1} + 0\imath$ 	& -  \\
Analytical,
direct fit 	& 	& $	5.0100931\times 10^{-1} -6.7261705\times 10^{-3}\imath$ 	& $4.9\times 10^{-3} + 6.7\times 10^{-3}\imath$\\
Analytical,
indirect fit & 	& $4.9608169\times 10^{-1} -3.0703754\times 10^{-6}\imath$	& $1.2\times 10^{-5} + 3.1\times 10^{-6}\imath$  \\
\hline
Numerical (reference) 	& 0.2 & $4.2082869\times 10^{-1} + 0\imath$ 	& -  \\
Analytical,
direct fit 	& 	& $1.8851877\times 10^{3} + 1.5091568\times 10^{3}\imath$ 	& $1.8\times 10^{3} + 1.5\times 10^{3}\imath$\\
Analytical,
indirect fit & 	& $4.1809499\times 10^{-1} -3.9984125\times 10^{-4}\imath$	& $2.7\times 10^{-3} + 4.0\times 10^{-4}\imath$  \\
\hline
\end{tabular}
}

\end{table}

A short numerical illustration of this formula is given in table \ref{tab:integral_comparison}.
We compare a fully numerical calculation of integral \eqref{eq:RadInteg_Idef_q0} taken as reference value and two results given by the cGTO expansions, either issuing from direct (eq. \eqref{eq:RadInteg_Jgauss}) or indirect fits (eq. \eqref{eq:integral_final}). We compare the results for more or less diffuse hypothetical values of the Slater exponent $\zeta$.
The direct fit leads to reasonable values of the selected integral only if $\zeta$ is not too small; the accuracy deteriorates with smaller $\zeta$ (meaning a larger extension of the integrand) since the direct fit is only valid within the fitting box. The integral obtained with the indirect fit is more stable and remains accurate even when it implies domains larger than the fitting box.
Choosing the indirect fitting method for representing continuum states with cGTOs thus guarantees that accurate results will be obtained whether the other states at play are well localized or not. 

\section{Conclusion }

We have shown that cGTOs can be made fully compatible with continuum radial functions by developing an optimization strategy in two directions.
We have explored the influence of research bounds on the optimal Gaussian exponents, leading to more flexible solutions and significantly better accuracies for cGTO representations of Coulomb radial functions.
We have then resolved the discrepancy between the oscillatory regime naturally present in every continuum function and the typical locality of cGTOs basis sets; this has been achieved by suggesting an approach based on partial factorization of the asymptotic oscillations followed by an indirect fitting of an appropriately defined distortion factor. The correct radial function can be easily reconstructed.
The computational cost of the indirect fitting is similar to the cost of direct fitting, but the new version is of better quality in the sense that exponents can span a more physical range and the asymptotic behaviour up to infinity can be guaranteed. 
 We have also shown that the indirect fitting maintains the analytical evaluation of useful one-electron integrals.
These results confirm the reliability of the cGTO expansion strategy applied to continuum states and will certainly improve the quality of further calculations in the context of molecular applications.
Before closing this paper, we would like to emphasize again the large advantage of performing electronic integrals through the evaluation of closed-form expressions. Even if the optimization step can cost several hours for a given set of radial wavefunctions, once the optimized cGTOs are known, they can be very efficiently used in the study of ionization problems \cite{ammar_2024}. 
Work is ongoing about the possible use of cGTO expansions for representing multicentric continuum wavefunctions and related applications.

\bibliographystyle{elsarticle-num}

\bibliography{AQC_vol91_leclerc}

\providecommand{\noopsort}[1]{}\providecommand{\singleletter}[1]{#1}%
\begin{thebibliography}{10}
\expandafter\ifx\csname url\endcsname\relax
  \def\url#1{\texttt{#1}}\fi
\expandafter\ifx\csname urlprefix\endcsname\relax\def\urlprefix{URL }\fi
\expandafter\ifx\csname href\endcsname\relax
  \def\href#1#2{#2} \def\path#1{#1}\fi

\bibitem{hill_2013}
J.~G. Hill, Gaussian basis sets for molecular applications, International
  Journal of Quantum Chemistry 113~(1) (2013) 21--34.

\bibitem{gill1994molecular}
P.~M. Gill, Molecular integrals over gaussian basis functions, in: Advances in
  quantum chemistry, Vol.~25, Elsevier, 1994, pp. 141--205.

\bibitem{shavitt1963}
I.~Shavitt, The {Gaussian} function in calculations of statistical mechanics
  and quantum mechanics., in: B.~Alder, S.~Fernbach, M.~Rotenberg (Eds.),
  Methods in Computational Physics, Academic Press, New York, 1963, pp. 1--45.

\bibitem{kaufmann_1989}
K.~Kaufmann, W.~Baumeister, M.~Jungen, Universal {Gaussian} basis sets for an
  optimum representation of {Rydberg} and continuum wavefunctions, Journal of
  Physics B: Atomic, Molecular and Optical Physics 22~(14) (1989) 2223.

\bibitem{Nestmann_1990}
B.~M. Nestmann, S.~D. Peyerimhoff, Optimized {Gaussian} basis sets for
  representation of continuum wavefunctions, Journal of Physics B: Atomic,
  Molecular and Optical Physics 23~(22) (1990) L773.

\bibitem{Faure_2002}
A.~Faure, J.~D. Gorfinkiel, L.~A. Morgan, J.~Tennyson, Gtobas: fitting
  continuum functions with {Gaussian}-type orbitals, Computer Physics
  Communications 144~(2) (2002) 224--241.

\bibitem{fiori_2012}
M.~Fiori, J.~E. Miraglia, New approach for approximating the continuum wave
  function by {Gaussian} basis set, Computer Physics Communications 183~(12)
  (2012) 2528--2534.

\bibitem{mavsin_2014}
Z.~Ma{\v{s}}{\'\i}n, J.~D. Gorfinkiel, Towards an accurate representation of
  the continuum in calculations of electron, positron and laser field
  interactions with molecules, Journal of Physics: Conference Series 490~(1)
  (2014) 012090.

\bibitem{marante_2014}
C.~Marante, L.~Argenti, F.~Mart{\'\i}n, Hybrid {Gaussian}--b-spline basis for
  the electronic continuum: Photoionization of atomic hydrogen, Physical Review
  A 90~(1) (2014) 012506.

\bibitem{zhu_2021}
Y.~Zhu, J.~M. Herbert, High harmonic spectra computed using time-dependent
  {Kohn-Sham} theory with {Gaussian} orbitals and a complex absorbing
  potential, The Journal of Chemical Physics 156~(20) (2022) 204123.

\bibitem{coccia_2021}
E.~Coccia, E.~Luppi, Time-dependent ab initio approaches for high-harmonic
  generation spectroscopy, Journal of Physics: Condensed Matter (2021).

\bibitem{wozniak_2021}
A.~P. Wo{\'z}niak, M.~Lesiuk, M.~Przybytek, D.~K. Efimov, J.~S.
  Prauzner-Bechcicki, M.~Mandrysz, M.~Ciappina, E.~Pisanty, J.~Zakrzewski,
  M.~Lewenstein, et~al., A systematic construction of {Gaussian} basis sets for
  the description of laser field ionization and high-harmonic generation, The
  Journal of Chemical Physics 154~(9) (2021) 094111.

\bibitem{witzorky_2021}
C.~Witzorky, G.~Paramonov, F.~Bouakline, R.~Jaquet, P.~Saalfrank, T.~Klamroth,
  {Gaussian}-type orbital calculations for high harmonic generation in
  vibrating molecules: Benchmarks for {H$_2^+$}, Journal of Chemical Theory and
  Computation 17~(12) (2021) 7353--7365.

\bibitem{gao_2021}
J.~W. Gao, T.~Miteva, Y.~Wu, J.~G. Wang, A.~Dubois, N.~Sisourat, Single-and
  double-ionization processes using {Gaussian}-type orbitals: Benchmark on
  antiproton-helium collisions in the kev-energy range, Physical Review A
  103~(3) (2021) L030803.

\bibitem{morassut_2022}
C.~Morassut, E.~Luppi, E.~Coccia, A {TD-CIS} study of high-harmonic generation
  of uracil cation fragments, Chemical Physics 559 (2022) 111515.

\bibitem{ammar_2020}
A.~Ammar, A.~Leclerc, L.~U. Ancarani, Fitting continuum wavefunctions with
  complex gaussians: Computation of ionization cross sections, Journal of
  Computational Chemistry 41~(27) (2020) 2365--2377.

\bibitem{ammar_2021_II}
A.~Ammar, A.~Leclerc, L.~U. Ancarani, Multicenter integrals involving complex
  gaussian-type functions, in: Advances in Quantum Chemistry, Vol.~83,
  Elsevier, 2021, pp. 287--304.

\bibitem{ammar_2021_I}
A.~Ammar, L.~U. Ancarani, A.~Leclerc, A complex gaussian approach to molecular
  photoionization, Journal of Computational Chemistry 42~(32) (2021)
  2294--2305.

\bibitem{ammar_2024}
A.~Ammar, A.~Leclerc, L.~U. Ancarani, Calculation of electron-impact and
  photoionization cross sections of methane using analytical gaussian
  integrals, Physical Review A 109~(5) (2024) 052810.

\bibitem{bateman1953}
H.~Bateman, A.~Erd\'elyi,
  \href{https://authors.library.caltech.edu/43491/}{Higher transcendental
  functions}, McGraw-Hill, New York, 1953.
\newline\urlprefix\url{https://authors.library.caltech.edu/43491/}

\bibitem{gradshteyn2014}
I.~S. Gradshteyn, I.~M. Ryzhik, Table of integrals, series, and products,
  Academic press, San Diego, 2014.

\bibitem{powell2009}
M.~J.~D. Powell, The bobyqa algorithm for bound constrained optimization
  without derivatives, Tech. rep., Department of Applied Mathematics and
  Theoretical Physics, Centre for Mathematical Sciences, Cambridge (2009).

\bibitem{ammar_2023}
A.~Ammar, A.~Leclerc, L.~U. Ancarani, On the use of complex gtos for the
  evaluation of radial integrals involving oscillating functions, in: Advances
  in Quantum Chemistry, Vol.~88, Elsevier, 2023, pp. 133--149.

\bibitem{salvat2019}
F.~Salvat, J.~M. Fern{\'a}ndez-Varea, {RADIAL}: a fortran subroutine package
  for the solution of the radial {Schr{\"o}dinger} and {Dirac} wave equations,
  Computer Physics Communications 240 (2019) 165--177.

\bibitem{abramowitz_1964}
M.~Abramowitz, {\relax I. A}.~Stegun, Handbook of Mathematical Functions with
  Formulas, Graphs, and Mathematical Tables, Dover, New York, 1964.

\bibitem{johansson2013}
F.~Johansson, et~al., mpmath: a {P}ython library for arbitrary-precision
  floating-point arithmetic (version 0.18), {\tt http://mpmath.org/} (December
  2013).

\bibitem{nardin1992}
M.~Nardin, W.~F. Perger, A.~Bhalla, Algorithm 707: Conhyp: A numerical
  evaluator of the confluent hypergeometric function for complex arguments of
  large magnitudes, ACM Transactions on Mathematical Software (TOMS) 18~(3)
  (1992) 345--349.

\end{thebibliography}

\end{document}